\documentclass[preprint,12pt,english]{elsarticle}

\usepackage{graphicx}
\usepackage{subfigure}
\usepackage{float}
\usepackage{color}
\usepackage[toc,page]{appendix}
\usepackage{graphicx}
\usepackage{lineno}
\usepackage{booktabs}

\usepackage{todonotes}
\usepackage{multirow}
\usepackage{url}
\journal{Nuclear Instrument and Method A}

\begin{document}

\begin{frontmatter}
\title{Background Evaluation for the Neutron Sources in the Daya Bay Experiment}

\author[1]{W.~Q.~Gu\corref{cor2}}
\author[3]{G.~F.~Cao}
\author[3]{X.~H.~Chen}
\author[2,6]{X.~P.~Ji}
\author[1]{G.~S.~Li}
\author[4,5,7]{J.~J.~Ling}
\author[1]{J.~Liu\corref{cor1}}
\author[4]{X.~Qian}
\author[5,8]{W.~Wang}

\address[1]{Department of Physics and Astronomy, Shanghai Jiao Tong University, Shanghai Laboratory for Particle Physics and Cosmology, Shanghai, China}
\address[2]{Department of Engineering Physics, Tsinghua University, Beijing, China}
\address[3]{Institute of High Energy Physics, Beijing, China}
\address[4]{Brookhaven National Laboratory, Upton, New York, USA}
\address[5]{Sun Yat-Sen (Zhongshan) University, Guangzhou, China}
\address[6]{School of Physics, Nankai University, Tianjin, China}
\address[7]{Department of Physics, University of Illinois at Urbana-Champaign, Urbana, Illinois, USA}
\address[8]{College of William and Mary, Williamsburg, Virginia, USA}

\cortext[cor2]{wenqiang.gu@sjtu.edu.cn}
\cortext[cor1]{jianglai.liu@sjtu.edu.cn}
\begin{abstract}

We present an evaluation of the background induced by $^{241}$Am-$^{13}$C 
neutron calibration sources in the Daya Bay reactor neutrino experiment.
As a significant background for electron-antineutrino detection at 
0.26$\pm$0.12 per detector per day on average, it has been estimated by 
a Monte Carlo simulation that was benchmarked by a special calibration data set.
This dedicated data set also provide the energy spectrum of the background.

\end{abstract}

\end{frontmatter}


\section{Introduction}
\label{sec:intro}
Neutrons are common source of background for most of the underground experiments. 
They can be produced hadronically by 
cosmic rays, or by ($\alpha$,n) and spontaneous fissions from environmental and internal
primordial radionuclides such as $^{238}$U~\cite{Formaggio:2004ge}. 
For these experiments, neutrons have many different ways to 
produce background, for example, 
by elastic and inelastic scatterings, nuclear activations, 
or nuclear captures~\cite{Ahmad:2002jz, Aprile:2013tov, Mei:2007zd, Agostini:2013tek, Abe:2008aa}. 

The underground Daya Bay neutrino experiment measures the neutrino oscillation driven 
by the mixing angle $\theta_{13}$
using the electron-antineutrinos from the Daya Bay nuclear power plant~\cite{An:2012eh}. 
The electron-antineutrinos are detected by the gadolinium-doped liquid scintillator (GdLS)
via the so-called inverse $\beta$-decay (IBD) reaction, 
$\bar\nu_e + p \rightarrow e^{+} + n$, producing a prompt positron signal and 
a delayed neutron capture signal on Gd with a total gamma energy of about 8~MeV.
Cosmic-ray induced neutrons are obvious background as the prompt recoil and delayed capture 
signals can fake the IBDs. 
In addition, there is another 
unique and more important neutron background
specific to the Daya Bay experiment
arising from the neutron calibration 
sources positioned close to the detector. In this paper, we 
provide an experimental evaluation of this background combining the results from 
a special calibration run with a Monte Carlo (MC) simulation.
The rest of this paper is organized as follows. The general description of this background and the 
two-component formalism is given in Sec.~\ref{sec:param}, followed by the evaluation of each component in 
Sec.~\ref{section:Rnlike} and Sec.~\ref{section:fE}. A detailed discussion on the
special calibration measurement will also be presented in Sec.~\ref{section:fE}.
Finally, in Sec.~\ref{sec:sum}, we conclude by summarizing the rate and energy 
spectrum of this background as well
as their uncertainties. 

\section{The Daya Bay experiment and the neutron source background}
\label{sec:param}
\begin{figure}[!htpb]
  \centering
  \includegraphics[width=0.5\textwidth]{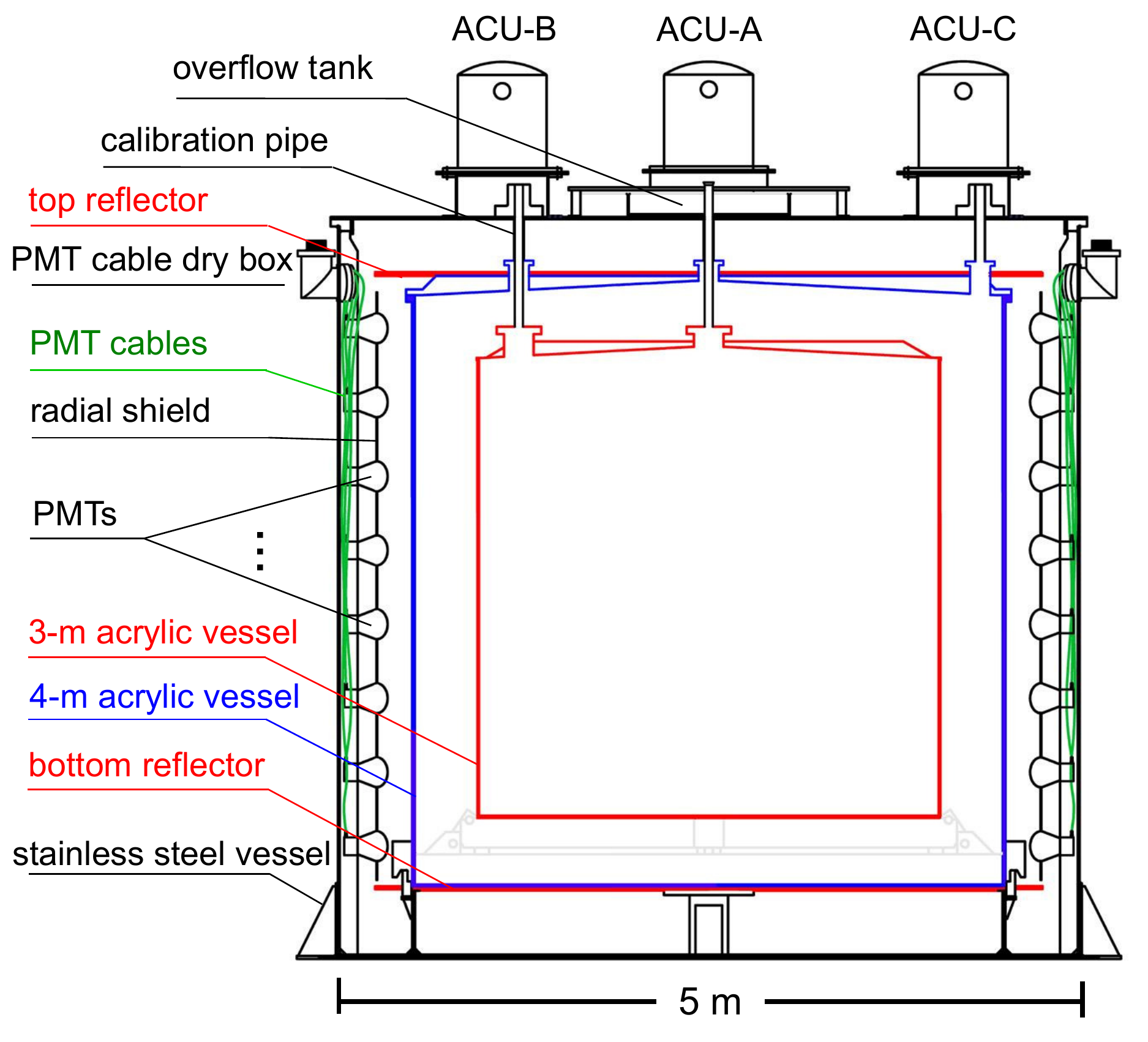}
  \caption{Side view of an AD with the three automatic calibration units (ACU-A, ACU-B and ACU-C) on the lid. Most of the enclosures are made out of SS. }
  \label{ACU_side}
\end{figure}

In the Daya Bay experiment, eight identical antineutrino detectors (ADs) are positioned 
in three experimental halls (two near halls and one far hall)~\cite{An:2015qga}. 
Two ADs are located in each near hall, 
and four ADs are positioned in the far hall near the $\theta_{13}$ oscillation maximum.
The antineutrino detector is built with three concentric cylindrical 
vessels as shown in Fig.~\ref{ACU_side}.
Approximately twenty tons of GdLS resides in the inner acrylic vessel.
The volume between the inner and outer acrylic vessels, known as the 
gamma catcher, is filled with un-doped liquid scintillator (LS) 
to improve the gamma detection 
efficiency. The volume between the outer acrylic vessel and the 24-ton stainless steel (SS, 
containing Fe: 70.8\%, Cr: 18\%, Ni: 8\%, Mn: 2\%, Si: 1\%,
C: 0.08\% in mass fractions) tank is filled with mineral oil to shield the ambient 
radiation as well as that from 
the photomultipliers (PMTs) and the SS tank.
The SS vessels are surrounded by two layers of water \v{C}erenkov detectors 
and resistive plate chambers which serve as the muon veto.

Three identical automated calibration units, ACU-A, B and C (Fig.~\ref{ACU_side}), 
located on the lid of each AD, deploy LED, gamma, and neutron sources 
vertically into the AD on regular basis to calibrate the detector response~\cite{Liu:2013ava}.
The sources are parked inside the ACU during regular data taking.
For the neutron source, traditional $^{252}$Cf and $^{241}$Am-$^{9}$Be sources have 
correlated multi-neutron and gamma-neutron emissions, respectively, which 
could lead to IBD-like background.
Therefore, specially designed $^{241}$Am-$^{13}$C (AmC) sources, each with
$\sim$0.7 neutron/s and free of correlated gamma-neutron emission, were used in 
Daya Bay~\cite{Liu:2015cra}. We shall refer to these sources as the 
LAS (the low-activity sources) hereafter.
According to a MC simulation with $10^8$ neutrons, none of these 
neutrons diffuses into the GdLS region 
where they would produce both a prompt recoil signal and a delayed capture signal on Gd.
However, being close to the AD all the time, a feeble but irreducible 
IBD-like background could still arise -- the prompt $\gamma$ 
produced by neutron inelastic scattering on the SS, followed by a 
high-energy gamma produced by the capture of the same neutron on 
the SS.
Such background will be referred to as the correlated background hereafter
\footnote{An alternative background caused by the random coincidence of two uncorrelated events,
the so called ``accidental background'', will be discussed in the Sec.~\ref{sec:uncertainty}.
It is not the main scope of this paper.
}.
To set the scale, each far site AD detects roughly 70 IBD reactions per day, 
and the neutron source background is about 0.2 events/day with an estimated uncertainty 
of 100\%~\cite{An:2012eh}. It was the most uncertain background in the Daya Bay
far site ADs. Due to its low rate, 
it is impractical to directly measure the correlated background by temporarily 
removing the neutron sources. 
On the other hand, the single SS capture signals from these source neutrons 
with reconstructed energy between 6 and 12 MeV has a rate of $\sim$230/day. They will
be referred to as the ``neutron-like'' events hereafter, as they mimic the 
single-neutron capture signals on Gd. 
We define
\begin{equation}
\label{eq:yield}
R_{\rm corr.} = R_{\rm neutron-like} \times \xi = R_{\rm neutron-like} \times \int_{E_{\rm min}}^{E_{\rm max}}f(E) dE\,
\end{equation}
where $R_{\rm corr.}$ and $R_{\rm neutron-like}$ are the rates of 
the correlated background and neutron-like events arising from the neutron sources, 
related by a ratio $\xi$. 
$f(E)$ is the differential 
form of $\xi$ as a function of the prompt energy, and $E_{\rm min}=0.7$~MeV and 
$E_{\rm max}=12$~MeV define the prompt-energy cut for the true IBDs. 
Since $R_{\rm neutron-like}$ can be directly measured and it is sensitive to the 
detector acceptance, Eq.~\ref{eq:yield} does not depend on the knowledge of
absolute source rate and allows at least 
partial cancellation of the systematic effects due to 
inaccuracy in the MC. 
For $\xi$ and $f(E)$, we performed direct
measurements by deploying a high-activity (but otherwise nearly identically designed) 
neutron source (HAS) on top of the detector and benchmarked them with the MC. 


\section{Neutron-like events: data and MC comparison}
\label{section:Rnlike}
The AmC background is studied with a Geant4-based~\cite{Agostinelli:2002hh} 
MC simulation (v4.9.2.p01) with detailed detector geometry. The neutron propagation and 
interaction is performed using the so-called High Precision 
neutron models~\cite{ref:g4manual}, 
which is largely based on the ENDF library~\cite{Chadwick:2011xwu}.
The so-called Low Energy electromagnetic processes are enabled for the gammas.
Neutron sources with their enclosure geometry are placed in the ACUs with expected energy spectrum implemented in the particle generator.
Realistic geometry of the detector, the ACU enclosure, as well as the
main interior components inside the ACUs are implemented in the MC.
Optical photons generated in the liquid scintillator are
tracked all the way until they hit the surface of the PMTs. 
In addition, the readout simulation is implemented according to 
the PMT response and the electronics model to convert optical photons to the 
charge collected by the readout electronics.
Based on the MC charge distribution on all PMTs, 
the vertex and energy are reconstructed with the same algorithm developed for 
data.

\begin{figure}[!htpb]
\centering
\includegraphics[width=0.75\textwidth]{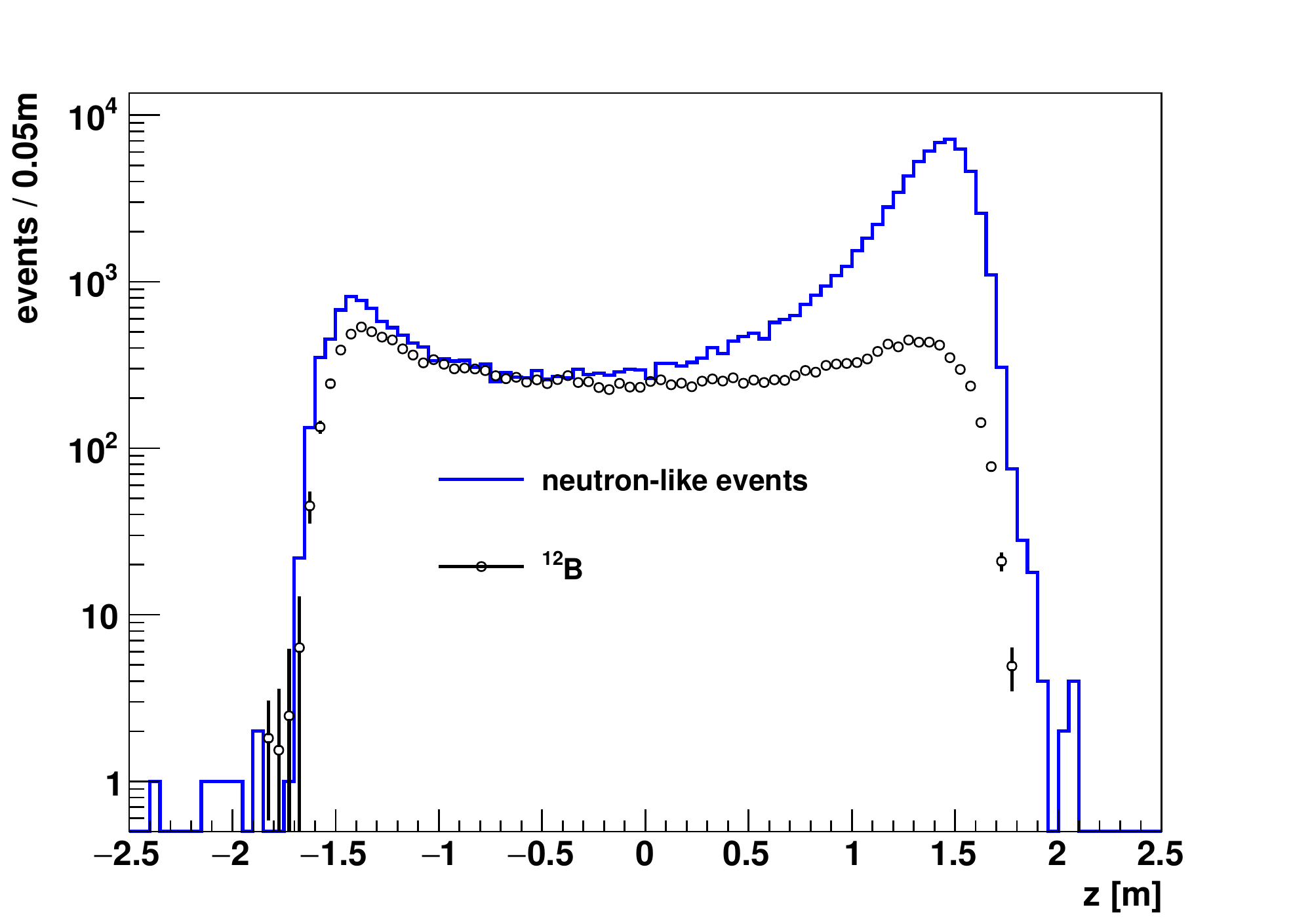}
\caption{Vertical distribution of the neutron-like events in a typical AD in the far site (blue histogram). 
The excess in upper half of the detector (Z$>$0) comes from the 
AmC sources in the ACUs. The distribution of tagged $^{12}$B events
are overlaid (black histogram) with an arbitrary vertical scale for visual clarity.
In both distributions, the excess of events close to $\pm$1.5~m 
is a position reconstruction artifact due to the top and bottom photon 
reflectors (see Fig.~\ref{ACU_side}).
}
\label{dataZdistro}
\end{figure}

As mentioned earlier, the neutron-like events from the AmC sources are produced by 
capture gammas from the SS elements.
In the data, we selected single neutron-like events that survive the 
muon veto from the AD and water \v{C}erenkov detectors.
Further details of muon veto will be provided in the Sec.~\ref{sec:uncertainty}.
The reconstructed vertical distribution of the neutron-like events
in a typical AD is shown in Fig.~\ref{dataZdistro}, where a strong 
excess from the top is observed.
Besides the AmC source, cosmogenic isotopes (e.g. $^{12}$B) 
that miss the muon veto also contribute to this distribution.
$^{12}$B is the major muon-induced $\beta$-emitting isotope with a 29.1~ms mean
lifetime. Due to the long lifetime, $^{12}$B events easily escape the muon veto,
and contribute to $>50\%$ of the non-AmC neutron-like events in all experimental halls.
This type of events is expected to distribute uniformly in the GdLS and LS
regions whereas the AmC induced background is expected to have the 
vertical distribution localized 
in the upper part of 
the AD. To confirm this, clean $^{12}$B 
events are selected within a 100~ms window after a showering muon 
(see Sec.~\ref{sec:uncertainty}). 
The vertical distribution of $^{12}$B
is overlaid in Fig.~\ref{dataZdistro}.
One observes good top-bottom symmetry with a difference less than 1.5\% 
combining all ADs. 

Based on the above, to extract the neutron-like events due to the AmC statistically, 
the difference between the 
top and bottom halves of the detector was taken. The average
$R_{\rm neutron-like}$ thus obtained is 230/day/AD,
with the values from individual detectors agreeing with each other within 30\%. 
For comparison, the top-bottom symmetric components (presumably non-AmC)  
are 498 and 48/day/AD in the near and far sites, respectively, with values 
from different detectors in agreement to better than 4\% in a given site. 
The 30\% AD-AD difference in the AmC component
could be due to the different absolute rates or spectra of the neutron sources, and we 
use it as a conservative estimate of the systematic uncertainty of $R_{\rm neutron-like}$.

\begin{figure}[!htpb]
\centering
\includegraphics[width=0.75\textwidth]{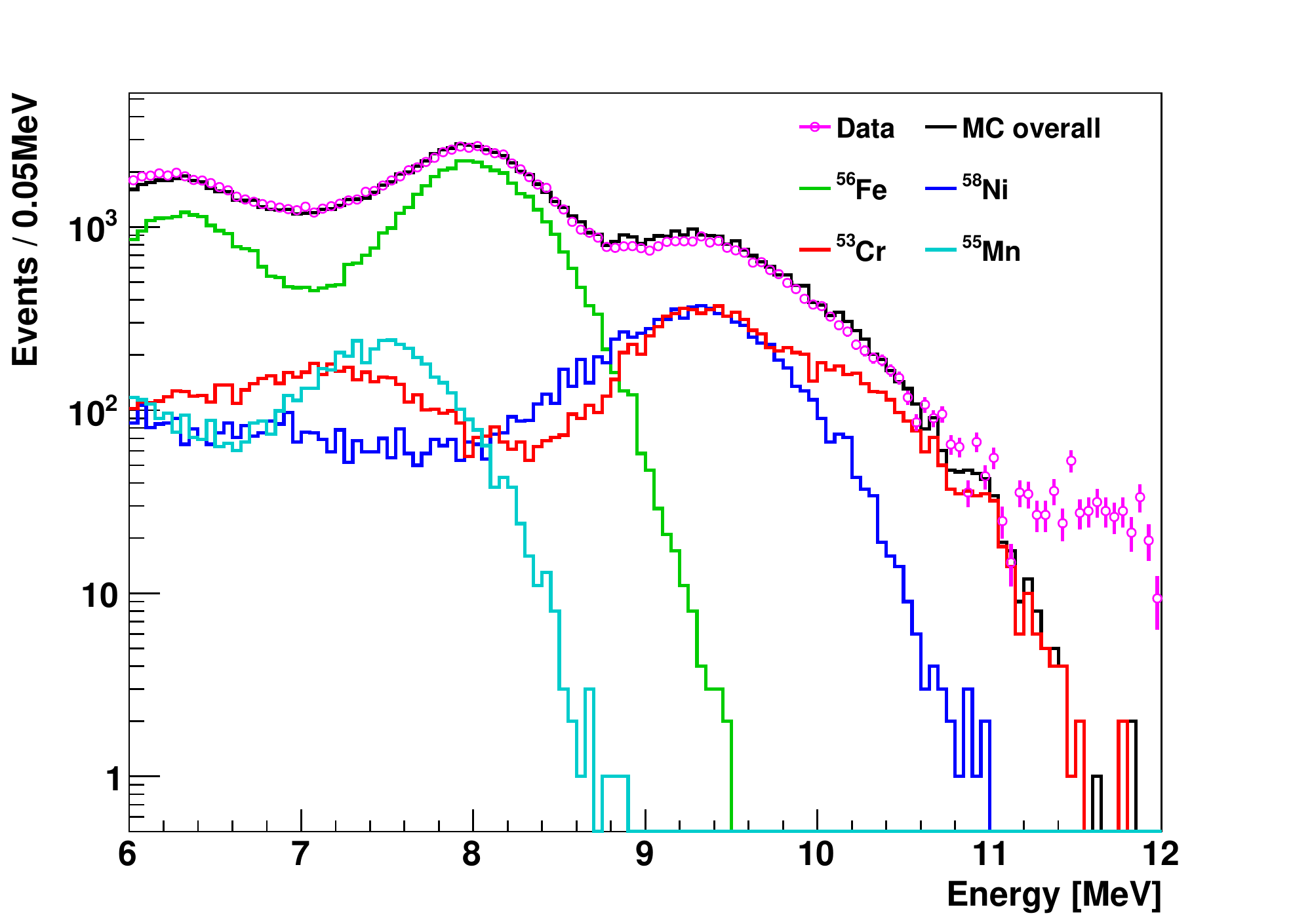}
\caption{Reconstructed energy spectrum for the LAS AmC 
neutron-like events: data versus MC.
Contributions from major neutron-capture targets from the MC are also shown.}
\label{singleComponent}
\end{figure}

The energy spectrum of the neutron-like events from the data 
is shown in Fig.~\ref{singleComponent}.
According to the MC, 
the main neutron-capture targets are $^{56}$Fe: 7.63~MeV, $^{58}$Ni: 9.00~MeV, 
$^{53}$Cr: 8.88~MeV, and $^{55}$Mn: 7.24~MeV. Their corresponding energy 
spectra are overlaid in Fig.~\ref{singleComponent},
which are in good agreement with data with no adjustment on the relative strength and spectrum of each composition.
The small excess above 11~MeV in the data,
likely due to an imperfect cancellation of the muon
induced products between the top and bottom,
is at a 0.5\% level to the total distribution and can be safely neglected.

\section{$\xi$ and $f(E)$: data and MC comparison}
\label{section:fE}
\subsection{Special calibration with HAS AmC}
\begin{figure}[!htpb]
	\centering
	\subfigure[HAS AmC source]{
	\label{fig:ss_enclosure}
	\includegraphics[width=0.35\textwidth]{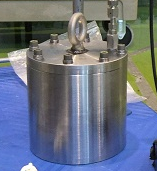}
	}
	\subfigure[Deployed onto the AD]{
	\label{strongPackAndLoc}
	\includegraphics[width=0.50\textwidth]{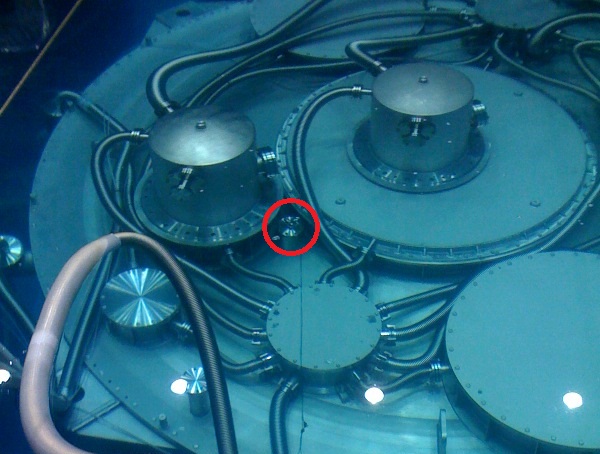}
      }
      \caption{(a) HAS AmC in a SS container; (b) picture of the HAS AmC
        deployed on the AD lid (red circle in the figure).
      }
    \end{figure}

To amplify the AmC background so that a direct measurement is possible, a 
HAS with 
$\sim$78 times intensity but otherwise nearly identical design as the LAS was 
prepared.
A special calibration run was performed with the HAS at the far site in Daya Bay
in the summer of 2012. To further amplify the rate of neutron interaction with the SS, 
the source was positioned at the center of a nearly solid cylindrical SS container
with 160~mm height and 165~mm diameter as shown in Fig.~\ref{fig:ss_enclosure}. 
The container was then deployed on the lid of an AD under water as shown in 
Fig.~\ref{strongPackAndLoc}.
Ten days of data were obtained with all ADs active and the water \v{C}erenkov 
detectors full but inactive (due to maintenance).
As the antineutrino detectors themselves are excellent muon detectors, 
the lack of active veto from the water \v{C}erenkov detectors is 
not important for the HAS data.

\subsection{Correlated background in HAS data}
\label{sec:uncertainty}
\begin{figure}[!htpb]
\centering
\includegraphics[scale=0.5]{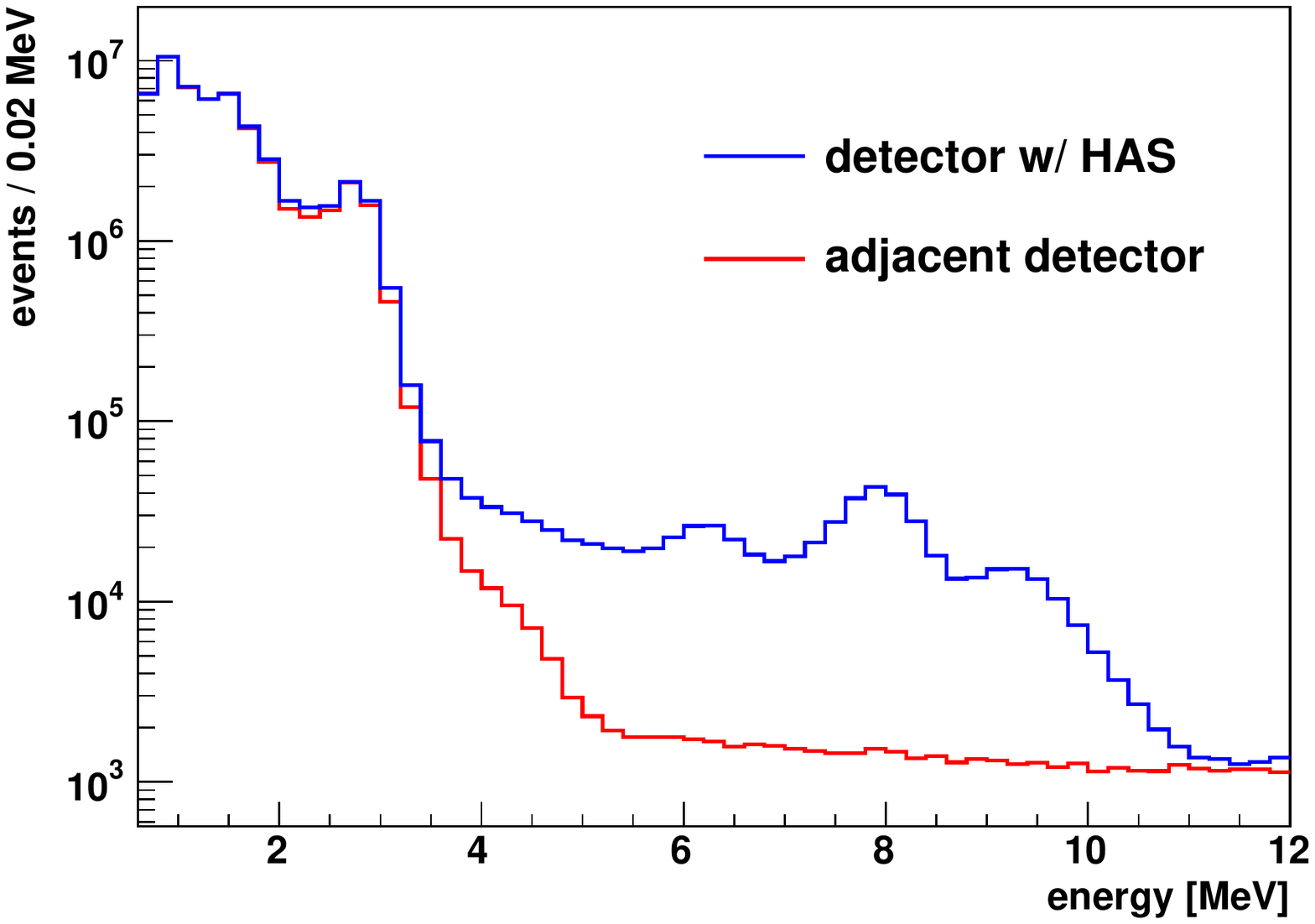}
\caption{Reconstructed neutron-like energy spectra 
in two neighboring detectors, one with the HAS 
on the lid (blue) and the other without (red).}
\label{strong_single_AD45}
\end{figure}

The event distributions solely due to the HAS can be 
obtained by taking the difference between the AD with the HAS (AD5) to an adjacent AD (AD4).
As an example, a comparison is made on the neutron-like energy spectra
in Fig.~\ref{strong_single_AD45}.
About 50,000 neutron-like events per day are observed 
in AD5 and only about 4,000 neutron-like per day are observed in AD4. 
The extra events in AD5 exhibit an expected 
energy spectrum from the neutron-capture gamma on the SS 
(see Fig.~\ref{singleComponent}).

Standard antineutrino event selection was made to extract
the correlated background from the high-activity AmC source. 
To avoid the contamination from the muon-induced background, each event 
in the detector is required to survive the AD muon veto~\cite{An:2012eh} -- ($-2$~$\mu$s, 
1000~$\mu$s) after the AD muon, and the showering-muon veto -- ($-2$~$\mu$s, 1~s) 
after the showering-muon. Here, the AD and showering-muon events are 
defined as any events with energy deposition $E>$20~MeV and $>$2.5~GeV (linear 
conversion from photoelectrons to energy) in the detector, respectively.
A time separation, (1~$\mu$s, 200~$\mu$s), between the prompt and delayed 
signals is required, and the energy of the prompt and delayed candidates are required 
to satisfy 0.7~MeV$ < E_{\rm prompt} < $12~MeV and 6~MeV$ < E_{\rm delayed} < $12~MeV, respectively.
There are not position cuts in the standard IBD selection in the Daya Bay experiment.

\begin{figure}[!htpb]
\centering
\includegraphics[width=0.75\textwidth]{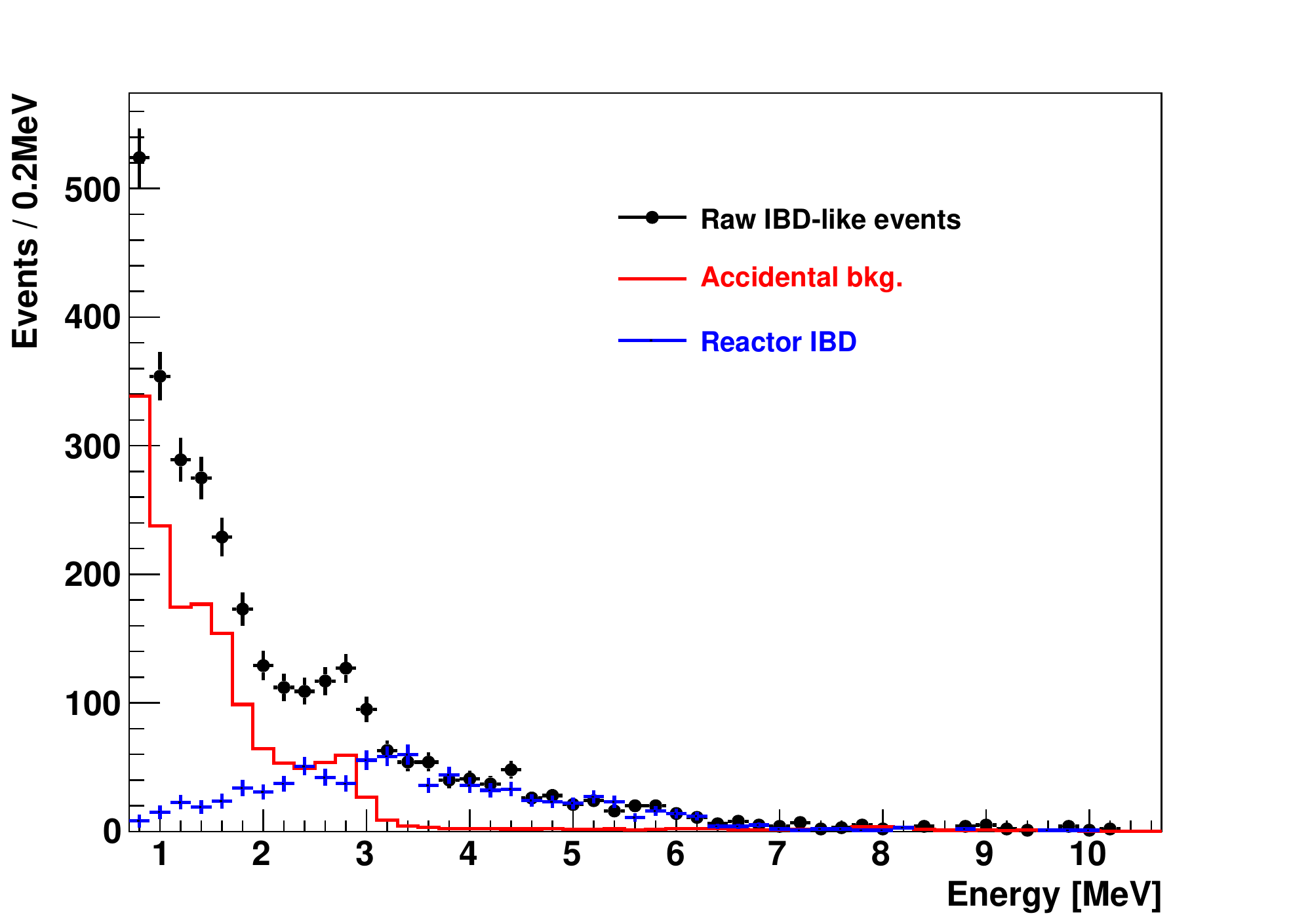}
\caption{The prompt energy spectra of raw IBD-like candidates from AD5 (black), 
the corresponding accidental background (red), and the reactor IBDs from AD4 (blue).
The accidental background spectrum exhibits the expected shape mainly from
natural radioactivities, and only 3\% of the accidental background
extends above 3.5~MeV. Compared to the uncertainty in the raw IBD-like events,
the uncertianty in the accidental background is negligible due to the large
statistics of single events.
}
\label{strong_accsub_prompt}
\end{figure}

The selected prompt-delayed pairs are contaminated by the coincidence
of two uncorrelated events, the so-called accidental background,
whose rate and spectrum can be precisely determined from the data by calculating
the probability that two signals randomly satisfy the requirement
of the IBD selection~\cite{An:2013uza}. In this case, ambient
radioactivity ($<$3.5 MeV) and the HAS neutron capture signals
dominate the prompt and delayed signals, respectively.
The prompt spectrum for the accidentals is predicted with that
from the single events (an isolated event in a coincidence time window)
without position cuts. Therefore, position dependence in the energy
spectrum is not relevant for the calculation of the accidental background.
For comparison, the prompt spectra of the 
raw IBD-like events from AD5, the corresponding accidental 
background, and the reactor IBDs obtained from AD4 are overlaid
in Fig.~\ref{strong_accsub_prompt}.
The HAS correlated background can then be obtained by subtracting the accidental background
and the reactor IBD from the raw IBD-like candidates,
as shown in Fig.~\ref{HAS_data_vs_MC}.

\subsection{Correlated background: from HAS to LAS}
\begin{figure}[!htpb]
	\centering
	\subfigure[HAS data vs. MC]{
	\label{HAS_data_vs_MC}
	\includegraphics[width=0.45\textwidth]{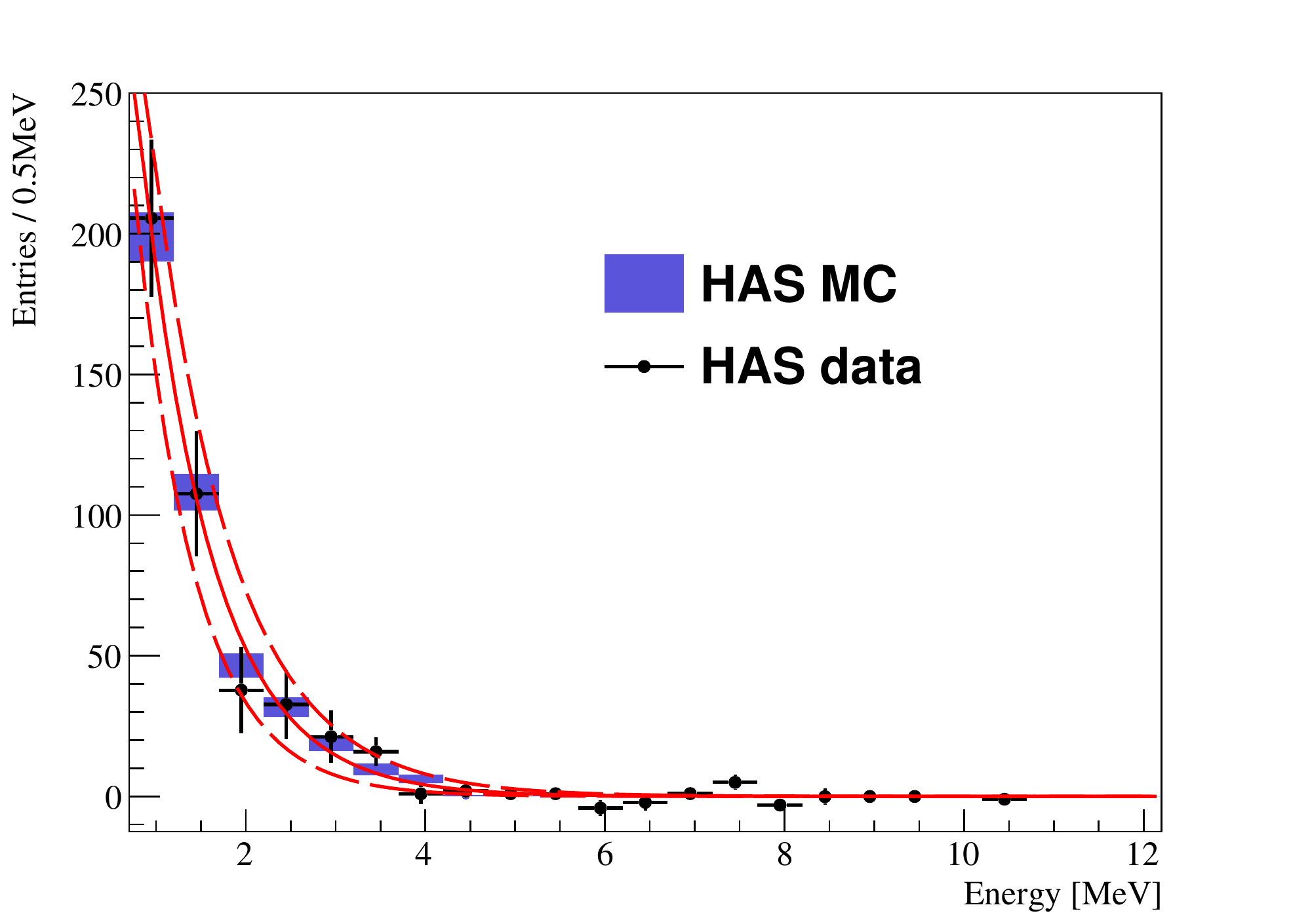}
	}
	\subfigure[HAS MC vs. LAS MC]{
	\label{HAS_vs_LAS_m13a}
	\includegraphics[width=0.45\textwidth]{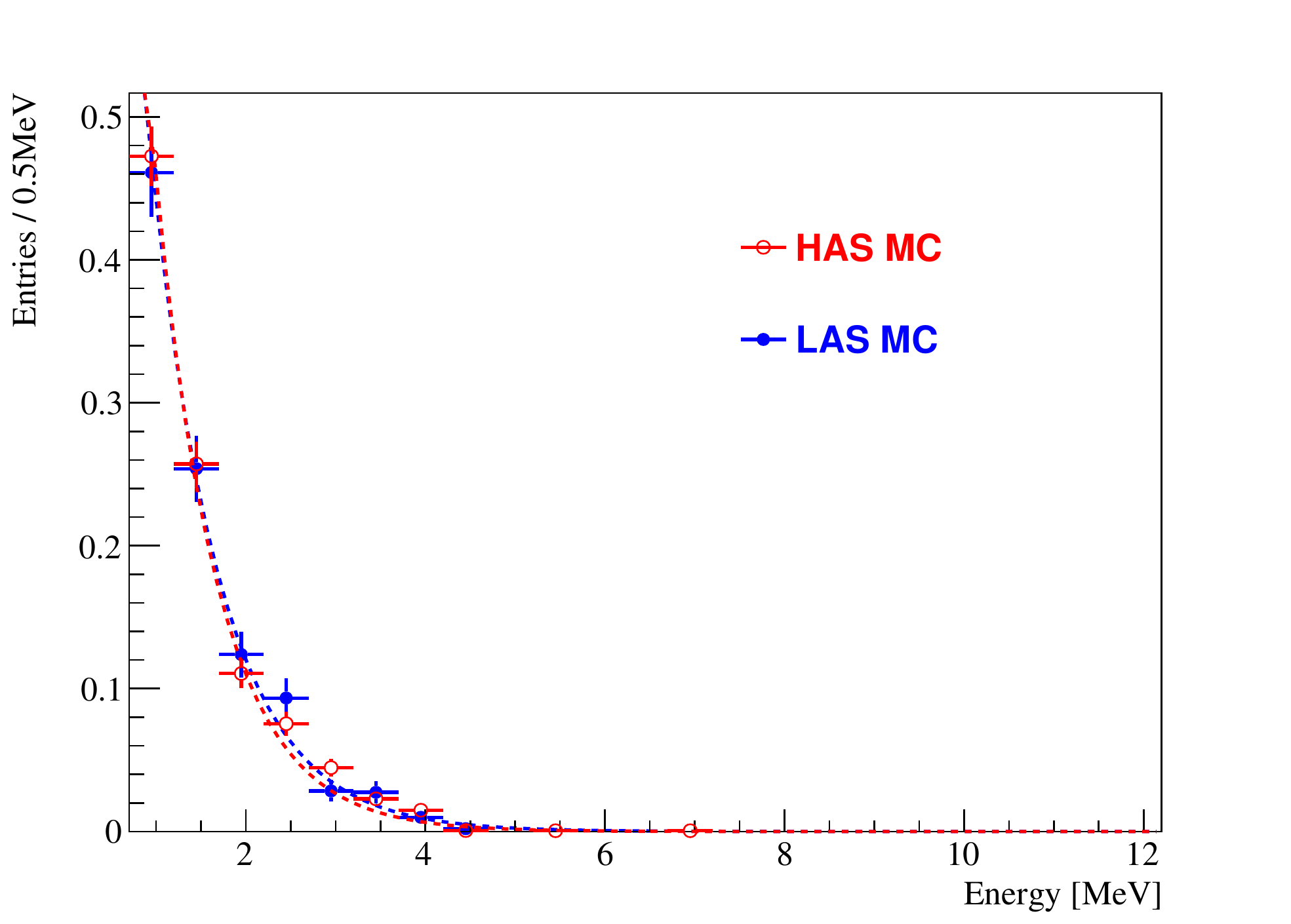}
      }
      \caption{Comparison of the prompt energy spectra for the correlated background. 
        (a) HAS data (black) vs. MC (blue with shaded error bars). 
        The uncertainties in data are mainly from statistics of the measured correlated
        events, while the uncertainty from the accidental background is negligible.
        An exponential fit is performed on the HAS data (red solid curve), 
        and the estimated uncertainty of this parameterization is encompassed by 
        the two red dashed curves. 
        The uncertainty of $p_1$ was estimated as 15\%, and its correlation
        with $p_0$ was conservatively ignored.
        (b) HAS MC (red) vs. LAS MC (blue), individually normalized. 
        Exponential fits are overlaid as the dashed curves.
      }
    \end{figure}

A MC simulation is performed for the HAS on the AD with its detailed geometry 
implemented. A two-stage comparison is then made on the data and MC. 
First, 
the comparison of the prompt-energy spectra between the data and MC
for the HAS correlated background is made in Fig.~\ref{HAS_data_vs_MC},
in which good agreement is found. 
Second, we compare the prompt spectra in the MC 
for the LAS and HAS correlated background in 
Fig.~\ref{HAS_vs_LAS_m13a}, and they also agree (see below 
for quantitative comparison on the shape). The two-step
agreement justifies that we can use the measured 
HAS spectrum to estimate the background due to the LAS.

The final measured HAS neutron-like and correlated background rates are 0.48 Hz and 63$\pm$13/day,
respectively, leading to a $\xi$ of (1.5$\pm$0.3)$\times10^{-3}$. The HAS
MC, in comparison, predicts 
a $\xi$ of (1.2$\pm$0.1)$\times10^{-3}$, and we attribute the 25\% difference 
as an uncertainty in the MC, including all systematic effects,
e.g. the difference of the MC geometry to the reality.
In combination with the 20\% statistical uncertainty from the data, we assign 
30\% as the total systematic uncertainty for $\xi$.
On the other hand, the predicted $\xi$ from the LAS MC is 
$0.9\times10^{-3}$, since the HAS is closer to the AD and with a thick SS housing by design.
According to the HAS data and MC comparison, we 
scale it up by 25\% to $\xi=1.125\times10^{-3}$ with an estimated 30\%
uncertainty. 
The prompt-energy spectrum of the 
correlated background is parameterized as an exponential function,
\begin{equation}
  \label{eq:fE}
  f(E) = p_0 \times e^{-E/p_{1}}\,,
\end{equation}
since the physical origin of the prompt energy is an accumulation of underlying 
Compton scatterings by the gamma rays produced by neutron inelastic scatterings 
on the SS.
The fit to the HAS data is shown in Fig.~\ref{HAS_data_vs_MC}.
The best fit value of $p_1$ is 0.783~MeV with 10\% statistical uncertainty,
and shifts very little when rebinning the data.
The fit functions, when varying $p_1$ by $\pm$15\%, are also shown in the figure,
which are sufficient to encompass possible shape variations in the data.
For comparison, $p_1$ obtained by fitting the LAS and HAS MC
distributions are 
0.794~MeV and 0.830~MeV (Fig.~\ref{HAS_vs_LAS_m13a}), 
respectively, both consistent to 6\% with that from the 
HAS data above.
We therefore assign 15\% as a conservative uncertainty for $p_1$ (uncorrelated to
$p_0$). 
The normalization condition for the LAS is given by Eq.~\ref{eq:yield} that  
$\xi = \int f(E) dE = 1.125 \times 10^{-3}$, from which we 
obtain $p_0 = 3.606\times10^{-3}$~/MeV, with a 30\% systematic 
uncertainty identical to that of $\xi$.
 


Given the common physical origin of the correlated background, 
the parameterization given by Eq.~\ref{eq:fE} 
and uncertainties are assumed to be global among the ADs.
In combination with $R_{\rm neutron-like}$ measured for each AD 
(Sec.~\ref{section:Rnlike}), the background rate
is evaluated to be 0.26$\pm$0.12 per AD per day. 
This improves the precision of the background evaluation significantly in comparison 
to that in Ref.~\cite{An:2012eh}.
Additionally, it provides the energy spectrum of the background, 
which is necessary for the spectral measurements of 
neutrino oscillation in Refs.~\cite{An:2013zwz} and \cite{An:2015rpe}.

\section{Summary and discussions}
\label{sec:sum}
A detailed study of the AmC neutron source induced background is performed with
extensive MC simulation, benchmarked by a calibration run with a special high-activity
AmC source at Daya Bay. 
The results are summarized in Table.~\ref{tabUncer}. 
On average the rate of the AmC background is 0.26 per AD per 
day with 45\% systematic uncertainty.
This confirms and improves the result in Ref.~\cite{An:2012eh} where an 
100\% uncertainty was assigned.
We parameterize the background with an exponential function 
$f(E) = p_0 \times e^{-E/p_{1}}$ where $p_0 = 3.606\times10^{-3}$~/MeV and 
$p_1=0.783$~MeV with 30\% and 15\% fractional uncertainties, respectively.
\begin{table}[!htpb]
  \caption{Summary of the mean values and 
    systematic uncertainties (in percentage to the mean) of the AmC background 
    used in the ``rate''~\cite{An:2012eh} and ``rate+shape''~\cite{An:2013zwz} 
    analysis of neutrino oscillation in Daya Bay. $R_{\rm IBD}$ in the table refers to the 
    average daily IBD rate in a far site detector.}
  \label{tabUncer}
  \centering
  \begin{tabular}{ccccc}
 &    \multicolumn{2}{c}{Rate Analysis} & \multicolumn{2}{c}{Rate$+$shape analysis}\\
     \cmidrule(l){1-3}\cmidrule(l){4-5}
$R_{\rm IBD}$ (/day) &     $R_{\rm neutron-like}$ (/day) & $\xi$ & $p_{0}$ (/MeV) & $p_{1}$ (MeV) \\
\multirow{2}{*}{70}  &   230 & 1.125$\times10^{-3}$ & 3.606$\times10^{-3}$& 0.783\\
  &   30\% & 30\% & 30\% & 15\% \\
    \bottomrule
  \end{tabular}
\end{table}

After seven months of operation at Daya Bay, through 
extensive calibration and data analysis, the relative detector 
responses were calibrated to high precision among ADs. It appeared 
that some AmC sources were no longer necessary at the far site. 
During the shutdown period in the summer of 2012 when the experiment completed the 
installation of the last two ADs, the AmC sources from ACU-B and C in 
all the far site ADs were removed. As a results, the AmC
background has been reduced by approximately a factor of three,
leading to a contribution of less than 0.1\% to the IBD signal rate, 
a major improvement highlighted in Ref.~\cite{An:2015rpe}. 
Further studies are ongoing to 
improve understandings of the AD-AD variation of the neutron-like rate, which currently
dominates the systematic uncertainty of the residual background.

The studies presented here is not only a crucial ingredient 
of the analysis of the Daya Bay data, but also provides potential benchmark data for 
neutron background in other reactor-based neutrino experiments.

\section*{Acknowledgement}
This work was done with support from the Natural Science Foundation of 
China grants 11175116, the Chinese Ministry of Science and Technology grant 
2013CB834306, Shanghai Laboratory for Particle Physics and Cosmology at 
the Shanghai Jiao Tong University, 
and the CAS Center for Excellence in Particle Physics (CCEPP).
We gratefully thank X.~L.~Chen from China Institute of Atomic Energy for the 
preparation of the high-activity $^{241}$Am-$^{13}$C source.

\bibliography{AmCBkg_NIM.bib}
\bibliographystyle{unsrt}

\end{document}